\documentclass[12pt]{article}
\usepackage{epsfig,a4}
\def\be{\begin{equation}}
\def\ee{\end{equation}}
\def\bea{\begin{eqnarray}}
\def\eea{\end{eqnarray}}
\def\rarr{\rightarrow}

\def\kf{{\bf k}}
\def\qf{{\bf q}}
\def\lf{{\bf l}}
\def\nn{\nonumber}
\def\fr{\frac}

\newdimen\picraise
\newcommand\picbox[1]
{
  \setbox0=\hbox{\input{#1}}
  \picraise=-0.5\ht0 
  \advance\picraise by 0.5\dp0
  \advance\picraise by 3pt     
  \hbox{\raise\picraise \box0}
}
\begin{document}
\begin{titlepage}
\begin{flushright}
Cavendish--HEP--99/10\\
DAMTP--1999--147\\
hep-ph/9911225
\end{flushright}
\vfill
\vspace*{1cm}
\begin{center}
\boldmath
{\Large{\bf New Representation for the 2-to-4 Gluon Vertex}}\\[.3cm]
{\Large{\bf in High Energy QCD $^*$}}
\unboldmath
\end{center}
\vspace{1.2cm}
\begin{center}
{\bf \large 
Carlo Ewerz
}
\end{center}
\vspace{.2cm}
\begin{center}
{\sl
Cavendish Laboratory, Cambridge University\\
Madingley Road, Cambridge CB3 0HE, UK\\[.2cm]}
and\\[.2cm]
{\sl 
DAMTP, Cambridge University\\ 
Silver Street, Cambridge CB3 9EW, UK\\[.4cm]}
email: {\sl carlo@hep.phy.cam.ac.uk}\\[.4cm]
\end{center}
\vfill
\begin{abstract}
A new representation for the 
two--to--four gluon vertex arising in the context of 
unitarity corrections is derived which involves only 
BFKL kernels. 
We discuss possible implications of this representation, 
including the possibility of finding the NLO corrections to the vertex. 
\end{abstract}
\vfill
\vspace{5em}
\hrule width 5.cm
\vspace*{.5em}
{\small \noindent 
$^*$Work supported in part by the EU Fourth Framework Programme
`Training and Mobility of Researchers', Network `Quantum Chromodynamics
and the Deep Structure of Elementary Particles',
contract FMRX-CT98-0194 (DG 12 - MIHT), and by 
the German Bundesministerium f\"ur 
Bildung, Wissenschaft, Forschung und Technologie.
}
\end{titlepage}

\section{Introduction}
\label{intro}

The scattering of small color dipoles is an important 
process which can be studied in perturbative QCD. 
At large energy, the leading logarithms of the energy can be resummed
resulting in the BFKL equation \cite{BFKL}. It describes 
the $t$-channel exchange of two gluons which are interacting via 
the so--called BFKL kernel. 
With increasing energy the parton densities become larger and parton 
recombination effects become important. In this situation subleading 
corrections to the BFKL equation have to be taken into account. 
These so--called unitarity corrections have recently been studied 
in the perturbative framework \cite{BW,BWL,BE}. 
They are characterized by exchanges with more than two reggeized 
gluons in the $t$-channel. Consequently, the quantities of interest in 
this context are amplitudes describing the production of $n$ gluons 
in the $t$-channel. 
Those $n$-gluon amplitudes have been found to exhibit interesting 
properties. 

The two most important properties of the $n$-gluon amplitudes 
are the field theory structure in these amplitudes and their conformal 
invariance in impact parameter space. The amplitudes with 
up to four gluons have been investigated in \cite{BW}. The three--gluon 
amplitude was shown to be a superposition of two--gluon (BFKL) 
amplitudes. The same is true for a part of the four--gluon amplitude, 
the so--called reggeizing part. The $t$-channel evolution of the other 
part of the four--gluon amplitude starts with 
a two--gluon state coupled to external particles. At some point during 
the $t$-channel evolution this two--gluon state undergoes a transition 
to a four--gluon state. The coupling of these two 
states is mediated by the two--to--four gluon vertex $V_{2\rarr 4}$. 
The vertex turns the set of quantum mechanical $n$-gluon states into 
a field theory. This field theory structure was shown to be present 
also in the five-- and six--gluon amplitudes \cite{BE}, indicating 
that the whole set of unitarity corrections can be formulated 
as an effective field theory. Also in those amplitudes the two--to--four 
vertex plays a prominent role. 
Further, the vertex $V_{2\rarr 4}$ was shown to be conformally 
invariant in two--dimensional 
impact parameter space \cite{BWL}. Together with the conformal 
invariance of the $n$-gluon states \cite{ngluonconf} this 
suggests that the effective field theory will be a conformal 
field theory. Recently, also the five-- and six--gluon amplitudes 
have been shown to be conformally invariant \cite{mythesis}. 

The two--to--four vertex is the simplest 
number--changing element and thus a key element for the 
understanding of the effective field theory. 
It is therefore important to study its properties and to 
gain deeper insight into its structure. 
In this letter we give a new representation of the vertex 
which we expect to be helpful in this respect. 
Surprisingly, it is possible to write the two--to--four vertex in 
terms of full BFKL kernels only. In other words, it can be 
expressed through simpler and well--understood elements. 
We are thus able to establish a relation between the 
vertex and the BFKL kernel. 
Although the new representation is rather long, it might 
be quite useful for the analysis of the vertex and its 
properties. We hope that it is also helpful for the comparison 
with other approaches to the unitarization problem like 
the dipole picture \cite{dipole} or the operator expansion 
\cite{operator}. As we will discuss, it might also allow one 
to find the NLO corrections to the two--to--four gluon vertex. 

The paper is organized as follows. In section \ref{BFKL} 
we recall some facts about the BFKL kernel and the two--to--four 
gluon vertex. In section \ref{newrep} we display and explain 
the new representation for this vertex. We also give a similar 
representation for another useful function arising in the 
analysis of unitarity corrections. Section \ref{implications} 
contains a discussion of the potential uses of the new representation 
and we close with a short summary. 

\section{The BFKL kernel and the two-to-four gluon vertex}
\label{BFKL}

The perturbative Pomeron is given by a two--gluon amplitude $D_2$ 
solving the BFKL equation
\be
\label{BFKLeq}
  \omega D_2 (\kf_1,\kf_2) = 
 D_{(2;0)}(\kf_1,\kf_2) 
 + \int \frac{d^2\lf}{(2\pi)^3} \,
 \frac{1}{\lf^2 (\qf-\lf)^2} 
 K_{\mbox{\scriptsize BFKL}}(\lf,\qf-\lf;\kf_1,\kf_2) 
\, D_2(\lf,\qf-\lf) 
\,,
\ee
where all momenta are two--dimensional transverse momenta, 
$\qf=\kf_1+\kf_2$ is the total momentum transfered in the $t$-channel, 
and $\omega$ is the complex angular momentum. 
$D_{(2;0)}(\kf_1,\kf_2)$ is an inhomogeneous term describing 
the coupling of the BFKL Pomeron to external particles, for 
example to virtual photons via a quark loop. 
The integral kernel in (\ref{BFKLeq}) is the BFKL kernel 
\bea
\label{Lipatovkernel}
K_{\mbox{\scriptsize BFKL}}(\lf,\qf-\lf;\kf,\qf-\kf) &=& 
 \! -  N_c g^2 \left[ \qf^2 - \frac{\kf^2(\qf-\lf)^2}{(\kf-\lf)^2} 
 - \frac{(\qf-\kf)^2 \lf^2}{(\kf-\lf)^2} \right] \nonumber \\
 &+& \! (2\pi)^3 \kf^2 (\qf-\kf)^2 
 \left[ \,\beta(\kf) + \beta(\qf - \kf) \right] 
 \delta(\kf-\lf).
\eea
The coupling constant $\alpha_s= g^2/(4\pi)$
is kept fixed, and the function $\beta$ is given by 
\be
\label{traject}
  \beta(\kf^2) = \frac{N_c}{2} g^2  \int \frac{d^2\lf}{(2 \pi)^3} 
          \frac{\kf^2}{\lf^2 (\lf -\kf)^2} 
\,.
\ee
The amplitudes $D_n$ describing the production of $n$ gluons 
in the $t$-channel are described by a set of coupled integral equations 
(for a review see \cite{BE}). The four--gluon amplitude 
for example satisfies
\be
  \left( \omega - \sum_{i=1}^4 \beta(\kf_i) \right) D_4 = 
D_{(4;0)}
+ K_{2\rarr 4} \otimes D_2 
+ \sum K_{2\rarr 3} \otimes D_3 
+ \sum K_{2\rarr 2} \otimes D_4
\label{inteq4}
\ee
where we have for simplicity suppressed all color indices. 
The convolution involves an integration like the one in (\ref{BFKLeq}), 
and the kernel $K_{2\rarr 2}$ is essentially given by the first 
square bracket in (\ref{Lipatovkernel}). 
The kernels $K_{2\rarr m}$ have been derived in 
\cite{Bartelskernels}. For a detailed description of the 
integral equation and the kernels we refer the reader to \cite{BE}. 
The three--gluon amplitude $D_3$ can be shown to 
reggeize, i.\,e.\ to be a superposition of two--gluon amplitudes. 
Then starting from eq.\ (\ref{inteq4}), the four--gluon amplitude can 
be shown to consist of 
two parts, $D_4=D_4^R + D_4^I$, the first of which (the reggeizing 
part) is a superposition 
of two--gluon amplitudes $D_2$. The second part, $D_4^I$, has the 
structure $D_4^I = G_4 \cdot V_{2 \rightarrow 4} \cdot D_2$, where 
$G_4$ is the Green's function of the four--gluon state. 
The vertex $V_{2 \rightarrow 4}$ thus couples the four--gluon to the 
two--gluon state. Its color structure is
\bea
\label{colV}
 V_{2 \rightarrow 4}^{a_1a_2a_3a_4}(\kf_1,\kf_2,\kf_3,\kf_4) 
 \!&=&\! 
  \delta_{a_1a_2} \delta_{a_3a_4} 
V(\kf_1,\kf_2;\kf_3,\kf_4)
 +\, \delta_{a_1a_3} \delta_{a_2a_4} 
V(\kf_1,\kf_3;\kf_2,\kf_4)
 \nonumber \\
 & &
 +\,\delta_{a_1a_4} \delta_{a_2a_3} 
V(\kf_1,\kf_4;\kf_2,\kf_3)
\,.
\eea
$V$ should be understood as an integral operator in momentum 
space, and when acting on $D_2$ its detailed form as derived 
in \cite{BW} is 
\bea
\label{v24}
\lefteqn{
(V D_2)  (\kf_1,\kf_2;\kf_3,\kf_4) =
\frac{g^4}{4} \times 
}
\nn \\
&\times&
\{\,2  \,[\, c(1234) 
-\, b(123,4) - b(124,3) - b(134,2) - b(234,1) 
\nn \\
&& 
\hspace{.2cm}
+ b(12,34) + b(34,12) 
+\, a(13,2,4) + a(14,2,3) + a(23,1,4) + a(24,1,3) \nn \\
&& 
\hspace{.2cm}
-\, a(1,2,34) - a(2,1,34) - a(3,12,4) - a(4,12,3) ]  \nn \\
&& 
+ [\, t(123,4) + t(124,3) + t(134,2) + t(234,1) -t(12,34) -t(34,12) 
\nn \\
&& 
\hspace{.2cm}
-\, s(13,2,4) - s(13,4,2) - s(14,2,3) - s(14,3,2) \nn \\
&& 
\hspace{.2cm}
-\, s(23,1,4) -s(23,4,1) - s(24,1,3) - s(24,3,1) \nn \\
&&  
\hspace{.2cm}
+\, s(1,2,34) + s(1,34,2) + s(2,1,34) + s(2,34,1) \nn \\
&& 
\hspace{.2cm}
+ \,s(3,12,4) + s(3,4,12)+ s(4,12,3) + s(4,3,12) ]  \} 
\,,
\eea
where the numbers in the arguments 
stand for the indices of the corresponding 
momenta $\kf_i$ and a string of numbers corresponds to 
the sum of the momenta. The functions in this expression are 
\bea
a(\kf_1,\kf_2,\kf_3) &=&  \int \fr{d^2\lf}{(2 \pi)^3}  
       \fr{\kf_1^2}{(\lf-\kf_2)^2 [\lf-(\kf_1+\kf_2)]^2}
      \,D_2\!\left(\lf, \sum_{j=1}^{3}\kf_j-\lf \right) \,,
\\
b(\kf_1,\kf_2) &=& a(\kf_1,\kf_2,\kf_3=0)\,, 
\\
c(\kf_1) &=& b(\kf_1,\kf_2=0)\,,
\\
s(\kf_1,\kf_2,\kf_3) &=& 
\fr{2}{N_c g^2} \beta(\kf_1)  D_2(\kf_1+\kf_2,\kf_3) \,,
\\
t(\kf_1,\kf_2) &=& s(\kf_1,\kf_3=0,\kf_2)
\eea
The first three of these correspond to real gluon emission, 
the last two describe virtual corrections. 

In \cite{BraunVacca} it was pointed out that the function $G$ 
introduced in \cite{BW}, 
\bea
\label{gdef}
 G(\kf_1,\kf_2,\kf_3) &=& \frac{g^2}{2}\, [
  \,2 c(123) - 2 b(12,3) - 2 b(23,1) + 2 a(2,1,3) \nn \\ 
 & &  \hspace{.6cm} +\, t(12,3) + t(23,1) - s(2,1,3) - s(2,3,1) ] 
\,,
\eea
is a very useful tool for studying the vertex $V_{2\rarr 4}$ because 
the vertex function $V$ can be represented as 
a superposition of $G$-functions, 
\bea
\label{vmitg}
  (V D_2 )(\kf_1,\kf_2;\kf_3,\kf_4) &=& \frac{g^2}{2} [ \, 
  G(1,23,4) + G(2,13,4) + G(1,24,3) + G(2,14,3) \nn \\ 
 & & \hspace{.5cm} 
  -\, G(12,3,4) - G(12,4,3) - G(1,2,34) - G(2,1,34) \nn \\
 & & \hspace{.5cm} +\, G(12,-,34) ] 
\,.
\eea
Already the simpler function $G$ is conformally invariant 
and infrared finite by itself. However, it   
does not vanish when its second argument vanishes, whereas 
the vertex function $V$ does have this property for all of its arguments. 
The function $G$ does not occur as an isolated object in the 
analysis of the integral equations, and only combinations similar 
to (\ref{vmitg}) are found \cite{mythesis}. 
We therefore consider the function $G$ to be of less fundamental 
significance than the full vertex function $V$. Nevertheless, it 
is a convenient object for computational purposes, for example 
for discussing the conformal invariance of the unitarity corrections 
\cite{BraunVacca,Vacca,mythesis}. 

\boldmath
\section{New representation for the vertex $V_{2\rarr 4}$}
\unboldmath
\label{newrep}

We now give a representation of the 
vertex function $V$ --- and thus of the full 
vertex $V_{2\rarr 4}$ --- which involves only BFKL kernels 
and free propagators ($1/\kf^2$). 

Let us outline how the new representation can be derived 
by reconsidering the usual derivation of $V$ (see \cite{BW,BE}). 
The basic idea is to express the kernels 
$K_{2\rarr m}$  with $m=3,4$ in the integral equation (\ref{inteq4}) 
in terms of two--to--two kernels $K_{2\rarr 2}$. 
(This was done for the two--to--three kernel $K_{2\rarr 3}$ 
also in \cite{Vacca,Braun}\footnote{In those references 
also the two--to--four kernel  $K_{2\rarr 4}$ 
was expressed in terms of the kernel $K_{2\rarr 2}$. 
The identity given there is not suited for deriving 
a new representation of the full vertex $V_{2 \rightarrow 4}$ 
in terms of full BFKL kernels 
as we give it here. We have found a different way 
to express $K_{2\rarr 4}$ in terms of $K_{2\rarr 2}$ 
which can be seen in the last square brackets of equation 
(\protect\ref{neudarst}) below.}.) 
Then the color tensors in the integral equation 
can be shown to arrange in such 
a way that the kernel $K_{2\rarr 2}$ can be 
replaced in these expressions by the full BFKL kernel 
by adding and subtracting appropriate trajectory functions $\beta$. 

To display the formula, we first define 
${\cal K}$ to be the product of a full BFKL kernel with the 
two propagators entering from above,
\be
{\cal K}(\qf_1,\qf_2;\kf_1,\kf_2) 
  = \fr{-1}{N_c g^2}
   \fr{1}{\qf_1^2} \fr{1}{\qf_2^2} 
     K_{\mbox{\scriptsize BFKL}}(\qf_1,\qf_2;\kf_1,\kf_2)
\,,
\label{kernelmitprop}
\ee
where (cf.\ (\ref{Lipatovkernel})) the kernel 
$K_{\mbox{\scriptsize BFKL}}$ includes the trajectory functions 
$\beta$. The vertex function $V$ 
(see (\ref{v24})) can then be written as
\bea
\label{neudarst}
\lefteqn{(V D_2) (\kf_1,\kf_2;\kf_3,\kf_4) = 
            - \fr{g^4}{4 (2 \pi )^3} \int 
            \left( \prod_{i=1}^{4} d^2\lf_i \right)
            \delta \left( \sum_{j=1}^4 \lf_j - \sum_{j=1}^4 \kf_j \right) 
\times
        } \nn \\
  &&\times  \Big\{ - \left[ D_2(\lf_1+\lf_2+\lf_3, \lf_4) 
                     + D_2(\lf_1+\lf_2+\lf_4,\lf_3)  
                     + D_2(\lf_1+\lf_3+\lf_4,\lf_2) 
                \right. \nn \\
  & &   \left. \hspace{1cm}
                   + D_2(\lf_1,\lf_2+\lf_3+\lf_4)
                   - D_2(\lf_1+\lf_2,\lf_3+\lf_4)
                     - D_2(\lf_1+\lf_3,\lf_2+\lf_4)
                \right. \nn \\
  & &   \left. \hspace{1cm}
                     - D_2(\lf_1+\lf_4,\lf_2+\lf_3)
                \right] 
                \nn \\
  & &  \hspace{.7cm}
              \times \left[ 
                 {\cal K}(\lf_1,\lf_2;\kf_1,\kf_2) 
                          \delta(\lf_3-\kf_3) \delta(\lf_4-\kf_4) 
                 \right. 
                 \nn \\
& &  \left. \hspace{1cm}
               + {\cal K} (\lf_3,\lf_4;\kf_3,\kf_4) 
                          \delta(\lf_1-\kf_1) \delta(\lf_2-\kf_2) 
              \right] 
              \nn \\
  & &   \hspace{.7cm}
                 + \left[ D_2(\lf_1+\lf_2+\lf_3, \lf_4) 
                      + D_2(\lf_1,\lf_2+\lf_3+\lf_4)
                      - D_2(\lf_1+\lf_4,\lf_2+\lf_3)
               \right]   
              \nn \\
  & &  \hspace{.7cm}
              \times \left[ 
                 {\cal K}(\lf_1,\lf_3;\kf_1,\kf_3) 
                          \delta(\lf_2-\kf_2) \delta(\lf_4-\kf_4) 
                  \right. 
                 \nn \\
& &  \left. \hspace{1cm}
               + {\cal K} (\lf_2,\lf_4;\kf_2,\kf_4) 
                          \delta(\lf_1-\kf_1) \delta(\lf_3-\kf_3) 
              \right] 
              \nn \\
  & &  \hspace{.7cm}
            + \left[ D_2(\lf_1+\lf_2+\lf_4, \lf_3) 
                     + D_2(\lf_1+\lf_3+\lf_4,\lf_2)
                     - D_2(\lf_1+\lf_2,\lf_3+\lf_4)
             \right.
             \nn \\
  & & \hspace{1cm}
             \left.
                     - D_2(\lf_1+\lf_3,\lf_2+\lf_4)
           \right]
              \nn \\
  & &  \hspace{.7cm}
              \times \left[ 
                 {\cal K}(\lf_1,\lf_4;\kf_1,\kf_4) 
                          \delta(\lf_2-\kf_2) \delta(\lf_3-\kf_3) 
                  \right. 
                 \nn \\
& &  \left. \hspace{1cm}
               + {\cal K} (\lf_2,\lf_3;\kf_2,\kf_3) 
                          \delta(\lf_1-\kf_1) \delta(\lf_4-\kf_4) 
              \right] 
              \Big\}  \nn \\
 &+& \fr{g^4}{4 (2 \pi )^3} \int 
            \left( \prod_{i=1}^{3} d^2\lf_i \right)
            \delta \left( \sum_{j=1}^3 \lf_j - \sum_{j=1}^4 \kf_j \right)
\times
           \nn \\
 & & 
         \times \left[ D_2(\lf_1+\lf_2,\lf_3) - D_2(\lf_1+\lf_3,\lf_2) 
                  + D_2(\lf_1,\lf_2+\lf_3) \right]  \nn \\
 &&\times \Big\{ 
          \left[ {\cal K}(\lf_1,\lf_2;\kf_1+\kf_2,\kf_3) 
                   -{\cal K}(\lf_1-\kf_1,\lf_2;\kf_2,\kf_3)
          \right] \delta(\lf_3-\kf_4) \nn \\
 & & \hspace{.8cm}
           - \left[ {\cal K}(\lf_1,\lf_3;\kf_1+\kf_2,\kf_4) 
                   -{\cal K}(\lf_1-\kf_1,\lf_3;\kf_2,\kf_4)
          \right] \delta(\lf_2-\kf_3) \nn \\
 & & \hspace{.8cm}
           - \left[ {\cal K}(\lf_1,\lf_3;\kf_1,\kf_3+\kf_4) 
                   -{\cal K}(\lf_1,\lf_3-\kf_4;\kf_1,\kf_3)
          \right] \delta(\lf_2-\kf_2) \nn \\
 & & \hspace{.8cm}
           +\left[ {\cal K}(\lf_2,\lf_3;\kf_2,\kf_3+\kf_4) 
                   -{\cal K}(\lf_2,\lf_3-\kf_4;\kf_2,\kf_3)
          \right] \delta(\lf_1-\kf_1) \Big\} \nn \\
 &+& \fr{g^4}{2 (2 \pi )^3} \int 
            \left( \prod_{i=1}^{2} d^2\lf_i \right)
            \delta \left( \sum_{j=1}^2 \lf_j - \sum_{j=1}^4 \kf_j \right)
            D_2(\lf_1,\lf_2) \times \nn \\
 &&\times \left[ {\cal K}(\lf_1,\lf_2;\kf_1+\kf_2,\kf_3+\kf_4) 
                  - {\cal K}(\lf_1-\kf_1,\lf_2;\kf_2,\kf_3+\kf_4)
         \right. \nn \\
 & & \hspace{.8cm}\left.           
                  - {\cal K}(\lf_1,\lf_2-\kf_4;\kf_1+\kf_2,\kf_3)
                  + {\cal K}(\lf_1-\kf_1,\lf_2-\kf_4;\kf_2,\kf_3) \right]
\eea
By introducing further $\delta$-functions for the arguments of 
the $D_2$'s one can easily isolate the vertex $V$ 
as an integral operator acting on $D_2$. 
The above representation can be shown to be equivalent 
to eq.\ (\ref{v24}) by a somewhat tedious but straightforward 
computation. 

Since eq.\ (\ref{neudarst}) is rather complicated we try to 
make it more transparent by using a diagrammatic notation. 
We define a diagram for ${\cal K}$, the BFKL kernel including the 
propagators for the gluons entering from above, 
\be
  {\cal K}(\qf_1,\qf_2;\kf_1,\kf_2) = \,\begin{picture}(0,0)%
\includegraphics{propkern.pstex}%
\end{picture}%
\setlength{\unitlength}{3947sp}%
\begingroup\makeatletter\ifx\SetFigFont\undefined
\def\x#1#2#3#4#5#6#7\relax{\def\x{#1#2#3#4#5#6}}%
\expandafter\x\fmtname xxxxxx\relax \def\y{splain}%
\ifx\x\y   
\gdef\SetFigFont#1#2#3{%
  \ifnum #1<17\tiny\else \ifnum #1<20\small\else
  \ifnum #1<24\normalsize\else \ifnum #1<29\large\else
  \ifnum #1<34\Large\else \ifnum #1<41\LARGE\else
     \huge\fi\fi\fi\fi\fi\fi
  \csname #3\endcsname}%
\else
\gdef\SetFigFont#1#2#3{\begingroup
  \count@#1\relax \ifnum 25<\count@\count@25\fi
  \def\x{\endgroup\@setsize\SetFigFont{#2pt}}%
  \expandafter\x
    \csname \romannumeral\the\count@ pt\expandafter\endcsname
    \csname @\romannumeral\the\count@ pt\endcsname
  \csname #3\endcsname}%
\fi
\fi\endgroup
\begin{picture}(330,657)(751,-211)
\put(751,-211){\makebox(0,0)[lb]{\smash{\SetFigFont{10}{12.0}{rm}$\kf_1$}}}
\put(1076,314){\makebox(0,0)[lb]{\smash{\SetFigFont{10}{12.0}{rm}$\qf_2$}}}
\put(751,314){\makebox(0,0)[lb]{\smash{\SetFigFont{10}{12.0}{rm}$\qf_1$}}}
\put(1081,-211){\makebox(0,0)[lb]{\smash{\SetFigFont{10}{12.0}{rm}$\kf_2$}}}
\end{picture}
 
\,.
\ee
Let us further introduce a pictorial notation for the momentum 
arguments of the BFKL amplitude $D_2$. We write 
\be
D_2(\lf_1+\lf_2,\lf_3+\lf_4) = 
D_2\left(\,\begin{picture}(0,0)%
\includegraphics{arg4_12.pstex}%
\end{picture}%
\setlength{\unitlength}{3947sp}%
\begingroup\makeatletter\ifx\SetFigFont\undefined
\def\x#1#2#3#4#5#6#7\relax{\def\x{#1#2#3#4#5#6}}%
\expandafter\x\fmtname xxxxxx\relax \def\y{splain}%
\ifx\x\y   
\gdef\SetFigFont#1#2#3{%
  \ifnum #1<17\tiny\else \ifnum #1<20\small\else
  \ifnum #1<24\normalsize\else \ifnum #1<29\large\else
  \ifnum #1<34\Large\else \ifnum #1<41\LARGE\else
     \huge\fi\fi\fi\fi\fi\fi
  \csname #3\endcsname}%
\else
\gdef\SetFigFont#1#2#3{\begingroup
  \count@#1\relax \ifnum 25<\count@\count@25\fi
  \def\x{\endgroup\@setsize\SetFigFont{#2pt}}%
  \expandafter\x
    \csname \romannumeral\the\count@ pt\expandafter\endcsname
    \csname @\romannumeral\the\count@ pt\endcsname
  \csname #3\endcsname}%
\fi
\fi\endgroup
\begin{picture}(399,249)(514,2)
\end{picture}
\right)
\,, 
\ee
and the generalization of the notation to other combinations 
of the four momenta $\lf_i$ is obvious. 
Now equation (\ref{neudarst}) can be rewritten as 
\bea
\lefteqn{\left(V D_2 \right) (\kf_1,\kf_2;\kf_3,\kf_4) = } \nn \\
    && - \fr{g^4}{4} 
     \bigg\{ - \left[ D_2\left(\,\begin{picture}(0,0)%
\includegraphics{arg4_123.pstex}%
\end{picture}%
\setlength{\unitlength}{3947sp}%
\begingroup\makeatletter\ifx\SetFigFont\undefined
\def\x#1#2#3#4#5#6#7\relax{\def\x{#1#2#3#4#5#6}}%
\expandafter\x\fmtname xxxxxx\relax \def\y{splain}%
\ifx\x\y   
\gdef\SetFigFont#1#2#3{%
  \ifnum #1<17\tiny\else \ifnum #1<20\small\else
  \ifnum #1<24\normalsize\else \ifnum #1<29\large\else
  \ifnum #1<34\Large\else \ifnum #1<41\LARGE\else
     \huge\fi\fi\fi\fi\fi\fi
  \csname #3\endcsname}%
\else
\gdef\SetFigFont#1#2#3{\begingroup
  \count@#1\relax \ifnum 25<\count@\count@25\fi
  \def\x{\endgroup\@setsize\SetFigFont{#2pt}}%
  \expandafter\x
    \csname \romannumeral\the\count@ pt\expandafter\endcsname
    \csname @\romannumeral\the\count@ pt\endcsname
  \csname #3\endcsname}%
\fi
\fi\endgroup
\begin{picture}(249,249)(514,2)
\end{picture}
\right) 
  + D_2\left(\,\begin{picture}(0,0)%
\includegraphics{arg4_124.pstex}%
\end{picture}%
\setlength{\unitlength}{3947sp}%
\begingroup\makeatletter\ifx\SetFigFont\undefined
\def\x#1#2#3#4#5#6#7\relax{\def\x{#1#2#3#4#5#6}}%
\expandafter\x\fmtname xxxxxx\relax \def\y{splain}%
\ifx\x\y   
\gdef\SetFigFont#1#2#3{%
  \ifnum #1<17\tiny\else \ifnum #1<20\small\else
  \ifnum #1<24\normalsize\else \ifnum #1<29\large\else
  \ifnum #1<34\Large\else \ifnum #1<41\LARGE\else
     \huge\fi\fi\fi\fi\fi\fi
  \csname #3\endcsname}%
\else
\gdef\SetFigFont#1#2#3{\begingroup
  \count@#1\relax \ifnum 25<\count@\count@25\fi
  \def\x{\endgroup\@setsize\SetFigFont{#2pt}}%
  \expandafter\x
    \csname \romannumeral\the\count@ pt\expandafter\endcsname
    \csname @\romannumeral\the\count@ pt\endcsname
  \csname #3\endcsname}%
\fi
\fi\endgroup
\begin{picture}(324,249)(514,2)
\end{picture}
\right) 
  + D_2\left(\,\begin{picture}(0,0)%
\includegraphics{arg4_134.pstex}%
\end{picture}%
\setlength{\unitlength}{3947sp}%
\begingroup\makeatletter\ifx\SetFigFont\undefined
\def\x#1#2#3#4#5#6#7\relax{\def\x{#1#2#3#4#5#6}}%
\expandafter\x\fmtname xxxxxx\relax \def\y{splain}%
\ifx\x\y   
\gdef\SetFigFont#1#2#3{%
  \ifnum #1<17\tiny\else \ifnum #1<20\small\else
  \ifnum #1<24\normalsize\else \ifnum #1<29\large\else
  \ifnum #1<34\Large\else \ifnum #1<41\LARGE\else
     \huge\fi\fi\fi\fi\fi\fi
  \csname #3\endcsname}%
\else
\gdef\SetFigFont#1#2#3{\begingroup
  \count@#1\relax \ifnum 25<\count@\count@25\fi
  \def\x{\endgroup\@setsize\SetFigFont{#2pt}}%
  \expandafter\x
    \csname \romannumeral\the\count@ pt\expandafter\endcsname
    \csname @\romannumeral\the\count@ pt\endcsname
  \csname #3\endcsname}%
\fi
\fi\endgroup
\begin{picture}(324,249)(214,2)
\end{picture}
\right) 
  + D_2\left(\,\,\begin{picture}(0,0)%
\includegraphics{arg4_234.pstex}%
\end{picture}%
\setlength{\unitlength}{3947sp}%
\begingroup\makeatletter\ifx\SetFigFont\undefined
\def\x#1#2#3#4#5#6#7\relax{\def\x{#1#2#3#4#5#6}}%
\expandafter\x\fmtname xxxxxx\relax \def\y{splain}%
\ifx\x\y   
\gdef\SetFigFont#1#2#3{%
  \ifnum #1<17\tiny\else \ifnum #1<20\small\else
  \ifnum #1<24\normalsize\else \ifnum #1<29\large\else
  \ifnum #1<34\Large\else \ifnum #1<41\LARGE\else
     \huge\fi\fi\fi\fi\fi\fi
  \csname #3\endcsname}%
\else
\gdef\SetFigFont#1#2#3{\begingroup
  \count@#1\relax \ifnum 25<\count@\count@25\fi
  \def\x{\endgroup\@setsize\SetFigFont{#2pt}}%
  \expandafter\x
    \csname \romannumeral\the\count@ pt\expandafter\endcsname
    \csname @\romannumeral\the\count@ pt\endcsname
  \csname #3\endcsname}%
\fi
\fi\endgroup
\begin{picture}(249,249)(439,2)
\end{picture}
\right) \right. \nn\\
  && \left.  \hspace{1.8cm}
   - D_2\left(\,\begin{picture}(0,0)%
\includegraphics{arg4_12.pstex}%
\end{picture}%
\setlength{\unitlength}{3947sp}%
\begingroup\makeatletter\ifx\SetFigFont\undefined
\def\x#1#2#3#4#5#6#7\relax{\def\x{#1#2#3#4#5#6}}%
\expandafter\x\fmtname xxxxxx\relax \def\y{splain}%
\ifx\x\y   
\gdef\SetFigFont#1#2#3{%
  \ifnum #1<17\tiny\else \ifnum #1<20\small\else
  \ifnum #1<24\normalsize\else \ifnum #1<29\large\else
  \ifnum #1<34\Large\else \ifnum #1<41\LARGE\else
     \huge\fi\fi\fi\fi\fi\fi
  \csname #3\endcsname}%
\else
\gdef\SetFigFont#1#2#3{\begingroup
  \count@#1\relax \ifnum 25<\count@\count@25\fi
  \def\x{\endgroup\@setsize\SetFigFont{#2pt}}%
  \expandafter\x
    \csname \romannumeral\the\count@ pt\expandafter\endcsname
    \csname @\romannumeral\the\count@ pt\endcsname
  \csname #3\endcsname}%
\fi
\fi\endgroup
\begin{picture}(399,249)(514,2)
\end{picture}
\right)
  - D_2\left(\,\begin{picture}(0,0)%
\includegraphics{arg4_13.pstex}%
\end{picture}%
\setlength{\unitlength}{3947sp}%
\begingroup\makeatletter\ifx\SetFigFont\undefined
\def\x#1#2#3#4#5#6#7\relax{\def\x{#1#2#3#4#5#6}}%
\expandafter\x\fmtname xxxxxx\relax \def\y{splain}%
\ifx\x\y   
\gdef\SetFigFont#1#2#3{%
  \ifnum #1<17\tiny\else \ifnum #1<20\small\else
  \ifnum #1<24\normalsize\else \ifnum #1<29\large\else
  \ifnum #1<34\Large\else \ifnum #1<41\LARGE\else
     \huge\fi\fi\fi\fi\fi\fi
  \csname #3\endcsname}%
\else
\gdef\SetFigFont#1#2#3{\begingroup
  \count@#1\relax \ifnum 25<\count@\count@25\fi
  \def\x{\endgroup\@setsize\SetFigFont{#2pt}}%
  \expandafter\x
    \csname \romannumeral\the\count@ pt\expandafter\endcsname
    \csname @\romannumeral\the\count@ pt\endcsname
  \csname #3\endcsname}%
\fi
\fi\endgroup
\begin{picture}(324,249)(514,2)
\end{picture}
\right)
  - D_2\left(\,\begin{picture}(0,0)%
\includegraphics{arg4_14.pstex}%
\end{picture}%
\setlength{\unitlength}{3947sp}%
\begingroup\makeatletter\ifx\SetFigFont\undefined
\def\x#1#2#3#4#5#6#7\relax{\def\x{#1#2#3#4#5#6}}%
\expandafter\x\fmtname xxxxxx\relax \def\y{splain}%
\ifx\x\y   
\gdef\SetFigFont#1#2#3{%
  \ifnum #1<17\tiny\else \ifnum #1<20\small\else
  \ifnum #1<24\normalsize\else \ifnum #1<29\large\else
  \ifnum #1<34\Large\else \ifnum #1<41\LARGE\else
     \huge\fi\fi\fi\fi\fi\fi
  \csname #3\endcsname}%
\else
\gdef\SetFigFont#1#2#3{\begingroup
  \count@#1\relax \ifnum 25<\count@\count@25\fi
  \def\x{\endgroup\@setsize\SetFigFont{#2pt}}%
  \expandafter\x
    \csname \romannumeral\the\count@ pt\expandafter\endcsname
    \csname @\romannumeral\the\count@ pt\endcsname
  \csname #3\endcsname}%
\fi
\fi\endgroup
\begin{picture}(307,249)(514,2)
\end{picture}
\right)\right]  
      \times \left( \,\,\begin{picture}(0,0)%
\includegraphics{ww12.pstex}%
\end{picture}%
\setlength{\unitlength}{3947sp}%
\begingroup\makeatletter\ifx\SetFigFont\undefined
\def\x#1#2#3#4#5#6#7\relax{\def\x{#1#2#3#4#5#6}}%
\expandafter\x\fmtname xxxxxx\relax \def\y{splain}%
\ifx\x\y   
\gdef\SetFigFont#1#2#3{%
  \ifnum #1<17\tiny\else \ifnum #1<20\small\else
  \ifnum #1<24\normalsize\else \ifnum #1<29\large\else
  \ifnum #1<34\Large\else \ifnum #1<41\LARGE\else
     \huge\fi\fi\fi\fi\fi\fi
  \csname #3\endcsname}%
\else
\gdef\SetFigFont#1#2#3{\begingroup
  \count@#1\relax \ifnum 25<\count@\count@25\fi
  \def\x{\endgroup\@setsize\SetFigFont{#2pt}}%
  \expandafter\x
    \csname \romannumeral\the\count@ pt\expandafter\endcsname
    \csname @\romannumeral\the\count@ pt\endcsname
  \csname #3\endcsname}%
\fi
\fi\endgroup
\begin{picture}(324,324)(889,-73)
\end{picture}

       + \,\begin{picture}(0,0)%
\includegraphics{ww34.pstex}%
\end{picture}%
\setlength{\unitlength}{3947sp}%
\begingroup\makeatletter\ifx\SetFigFont\undefined
\def\x#1#2#3#4#5#6#7\relax{\def\x{#1#2#3#4#5#6}}%
\expandafter\x\fmtname xxxxxx\relax \def\y{splain}%
\ifx\x\y   
\gdef\SetFigFont#1#2#3{%
  \ifnum #1<17\tiny\else \ifnum #1<20\small\else
  \ifnum #1<24\normalsize\else \ifnum #1<29\large\else
  \ifnum #1<34\Large\else \ifnum #1<41\LARGE\else
     \huge\fi\fi\fi\fi\fi\fi
  \csname #3\endcsname}%
\else
\gdef\SetFigFont#1#2#3{\begingroup
  \count@#1\relax \ifnum 25<\count@\count@25\fi
  \def\x{\endgroup\@setsize\SetFigFont{#2pt}}%
  \expandafter\x
    \csname \romannumeral\the\count@ pt\expandafter\endcsname
    \csname @\romannumeral\the\count@ pt\endcsname
  \csname #3\endcsname}%
\fi
\fi\endgroup
\begin{picture}(324,324)(739,-73)
\end{picture}
 \,\right)  \nn \\
  && \hspace{1.1cm} 
     +  \left[ D_2\left(\,\begin{picture}(0,0)%
\includegraphics{arg4_123.pstex}%
\end{picture}%
\setlength{\unitlength}{3947sp}%
\begingroup\makeatletter\ifx\SetFigFont\undefined
\def\x#1#2#3#4#5#6#7\relax{\def\x{#1#2#3#4#5#6}}%
\expandafter\x\fmtname xxxxxx\relax \def\y{splain}%
\ifx\x\y   
\gdef\SetFigFont#1#2#3{%
  \ifnum #1<17\tiny\else \ifnum #1<20\small\else
  \ifnum #1<24\normalsize\else \ifnum #1<29\large\else
  \ifnum #1<34\Large\else \ifnum #1<41\LARGE\else
     \huge\fi\fi\fi\fi\fi\fi
  \csname #3\endcsname}%
\else
\gdef\SetFigFont#1#2#3{\begingroup
  \count@#1\relax \ifnum 25<\count@\count@25\fi
  \def\x{\endgroup\@setsize\SetFigFont{#2pt}}%
  \expandafter\x
    \csname \romannumeral\the\count@ pt\expandafter\endcsname
    \csname @\romannumeral\the\count@ pt\endcsname
  \csname #3\endcsname}%
\fi
\fi\endgroup
\begin{picture}(249,249)(514,2)
\end{picture}
\right) 
    + D_2\left(\,\,\begin{picture}(0,0)%
\includegraphics{arg4_234.pstex}%
\end{picture}%
\setlength{\unitlength}{3947sp}%
\begingroup\makeatletter\ifx\SetFigFont\undefined
\def\x#1#2#3#4#5#6#7\relax{\def\x{#1#2#3#4#5#6}}%
\expandafter\x\fmtname xxxxxx\relax \def\y{splain}%
\ifx\x\y   
\gdef\SetFigFont#1#2#3{%
  \ifnum #1<17\tiny\else \ifnum #1<20\small\else
  \ifnum #1<24\normalsize\else \ifnum #1<29\large\else
  \ifnum #1<34\Large\else \ifnum #1<41\LARGE\else
     \huge\fi\fi\fi\fi\fi\fi
  \csname #3\endcsname}%
\else
\gdef\SetFigFont#1#2#3{\begingroup
  \count@#1\relax \ifnum 25<\count@\count@25\fi
  \def\x{\endgroup\@setsize\SetFigFont{#2pt}}%
  \expandafter\x
    \csname \romannumeral\the\count@ pt\expandafter\endcsname
    \csname @\romannumeral\the\count@ pt\endcsname
  \csname #3\endcsname}%
\fi
\fi\endgroup
\begin{picture}(249,249)(439,2)
\end{picture}
\right) 
    - D_2\left(\,\begin{picture}(0,0)%
\includegraphics{arg4_14.pstex}%
\end{picture}%
\setlength{\unitlength}{3947sp}%
\begingroup\makeatletter\ifx\SetFigFont\undefined
\def\x#1#2#3#4#5#6#7\relax{\def\x{#1#2#3#4#5#6}}%
\expandafter\x\fmtname xxxxxx\relax \def\y{splain}%
\ifx\x\y   
\gdef\SetFigFont#1#2#3{%
  \ifnum #1<17\tiny\else \ifnum #1<20\small\else
  \ifnum #1<24\normalsize\else \ifnum #1<29\large\else
  \ifnum #1<34\Large\else \ifnum #1<41\LARGE\else
     \huge\fi\fi\fi\fi\fi\fi
  \csname #3\endcsname}%
\else
\gdef\SetFigFont#1#2#3{\begingroup
  \count@#1\relax \ifnum 25<\count@\count@25\fi
  \def\x{\endgroup\@setsize\SetFigFont{#2pt}}%
  \expandafter\x
    \csname \romannumeral\the\count@ pt\expandafter\endcsname
    \csname @\romannumeral\the\count@ pt\endcsname
  \csname #3\endcsname}%
\fi
\fi\endgroup
\begin{picture}(307,249)(514,2)
\end{picture}
\right) \right] 
    \times \left(\,\,\begin{picture}(0,0)%
\includegraphics{ww13.pstex}%
\end{picture}%
\setlength{\unitlength}{3947sp}%
\begingroup\makeatletter\ifx\SetFigFont\undefined
\def\x#1#2#3#4#5#6#7\relax{\def\x{#1#2#3#4#5#6}}%
\expandafter\x\fmtname xxxxxx\relax \def\y{splain}%
\ifx\x\y   
\gdef\SetFigFont#1#2#3{%
  \ifnum #1<17\tiny\else \ifnum #1<20\small\else
  \ifnum #1<24\normalsize\else \ifnum #1<29\large\else
  \ifnum #1<34\Large\else \ifnum #1<41\LARGE\else
     \huge\fi\fi\fi\fi\fi\fi
  \csname #3\endcsname}%
\else
\gdef\SetFigFont#1#2#3{\begingroup
  \count@#1\relax \ifnum 25<\count@\count@25\fi
  \def\x{\endgroup\@setsize\SetFigFont{#2pt}}%
  \expandafter\x
    \csname \romannumeral\the\count@ pt\expandafter\endcsname
    \csname @\romannumeral\the\count@ pt\endcsname
  \csname #3\endcsname}%
\fi
\fi\endgroup
\begin{picture}(324,324)(889,-73)
\end{picture}
 
      + \,\begin{picture}(0,0)%
\includegraphics{ww24.pstex}%
\end{picture}%
\setlength{\unitlength}{3947sp}%
\begingroup\makeatletter\ifx\SetFigFont\undefined
\def\x#1#2#3#4#5#6#7\relax{\def\x{#1#2#3#4#5#6}}%
\expandafter\x\fmtname xxxxxx\relax \def\y{splain}%
\ifx\x\y   
\gdef\SetFigFont#1#2#3{%
  \ifnum #1<17\tiny\else \ifnum #1<20\small\else
  \ifnum #1<24\normalsize\else \ifnum #1<29\large\else
  \ifnum #1<34\Large\else \ifnum #1<41\LARGE\else
     \huge\fi\fi\fi\fi\fi\fi
  \csname #3\endcsname}%
\else
\gdef\SetFigFont#1#2#3{\begingroup
  \count@#1\relax \ifnum 25<\count@\count@25\fi
  \def\x{\endgroup\@setsize\SetFigFont{#2pt}}%
  \expandafter\x
    \csname \romannumeral\the\count@ pt\expandafter\endcsname
    \csname @\romannumeral\the\count@ pt\endcsname
  \csname #3\endcsname}%
\fi
\fi\endgroup
\begin{picture}(324,324)(814,-73)
\end{picture}
 \,\right) \nn \\
  && \hspace{1.1cm} 
    +  \left[  D_2\left(\,\begin{picture}(0,0)%
\includegraphics{arg4_124.pstex}%
\end{picture}%
\setlength{\unitlength}{3947sp}%
\begingroup\makeatletter\ifx\SetFigFont\undefined
\def\x#1#2#3#4#5#6#7\relax{\def\x{#1#2#3#4#5#6}}%
\expandafter\x\fmtname xxxxxx\relax \def\y{splain}%
\ifx\x\y   
\gdef\SetFigFont#1#2#3{%
  \ifnum #1<17\tiny\else \ifnum #1<20\small\else
  \ifnum #1<24\normalsize\else \ifnum #1<29\large\else
  \ifnum #1<34\Large\else \ifnum #1<41\LARGE\else
     \huge\fi\fi\fi\fi\fi\fi
  \csname #3\endcsname}%
\else
\gdef\SetFigFont#1#2#3{\begingroup
  \count@#1\relax \ifnum 25<\count@\count@25\fi
  \def\x{\endgroup\@setsize\SetFigFont{#2pt}}%
  \expandafter\x
    \csname \romannumeral\the\count@ pt\expandafter\endcsname
    \csname @\romannumeral\the\count@ pt\endcsname
  \csname #3\endcsname}%
\fi
\fi\endgroup
\begin{picture}(324,249)(514,2)
\end{picture}
\right) 
    + D_2\left(\,\begin{picture}(0,0)%
\includegraphics{arg4_134.pstex}%
\end{picture}%
\setlength{\unitlength}{3947sp}%
\begingroup\makeatletter\ifx\SetFigFont\undefined
\def\x#1#2#3#4#5#6#7\relax{\def\x{#1#2#3#4#5#6}}%
\expandafter\x\fmtname xxxxxx\relax \def\y{splain}%
\ifx\x\y   
\gdef\SetFigFont#1#2#3{%
  \ifnum #1<17\tiny\else \ifnum #1<20\small\else
  \ifnum #1<24\normalsize\else \ifnum #1<29\large\else
  \ifnum #1<34\Large\else \ifnum #1<41\LARGE\else
     \huge\fi\fi\fi\fi\fi\fi
  \csname #3\endcsname}%
\else
\gdef\SetFigFont#1#2#3{\begingroup
  \count@#1\relax \ifnum 25<\count@\count@25\fi
  \def\x{\endgroup\@setsize\SetFigFont{#2pt}}%
  \expandafter\x
    \csname \romannumeral\the\count@ pt\expandafter\endcsname
    \csname @\romannumeral\the\count@ pt\endcsname
  \csname #3\endcsname}%
\fi
\fi\endgroup
\begin{picture}(324,249)(214,2)
\end{picture}
\right) 
   - D_2\left(\,\begin{picture}(0,0)%
\includegraphics{arg4_12.pstex}%
\end{picture}%
\setlength{\unitlength}{3947sp}%
\begingroup\makeatletter\ifx\SetFigFont\undefined
\def\x#1#2#3#4#5#6#7\relax{\def\x{#1#2#3#4#5#6}}%
\expandafter\x\fmtname xxxxxx\relax \def\y{splain}%
\ifx\x\y   
\gdef\SetFigFont#1#2#3{%
  \ifnum #1<17\tiny\else \ifnum #1<20\small\else
  \ifnum #1<24\normalsize\else \ifnum #1<29\large\else
  \ifnum #1<34\Large\else \ifnum #1<41\LARGE\else
     \huge\fi\fi\fi\fi\fi\fi
  \csname #3\endcsname}%
\else
\gdef\SetFigFont#1#2#3{\begingroup
  \count@#1\relax \ifnum 25<\count@\count@25\fi
  \def\x{\endgroup\@setsize\SetFigFont{#2pt}}%
  \expandafter\x
    \csname \romannumeral\the\count@ pt\expandafter\endcsname
    \csname @\romannumeral\the\count@ pt\endcsname
  \csname #3\endcsname}%
\fi
\fi\endgroup
\begin{picture}(399,249)(514,2)
\end{picture}
\right)
   - D_2\left(\,\begin{picture}(0,0)%
\includegraphics{arg4_13.pstex}%
\end{picture}%
\setlength{\unitlength}{3947sp}%
\begingroup\makeatletter\ifx\SetFigFont\undefined
\def\x#1#2#3#4#5#6#7\relax{\def\x{#1#2#3#4#5#6}}%
\expandafter\x\fmtname xxxxxx\relax \def\y{splain}%
\ifx\x\y   
\gdef\SetFigFont#1#2#3{%
  \ifnum #1<17\tiny\else \ifnum #1<20\small\else
  \ifnum #1<24\normalsize\else \ifnum #1<29\large\else
  \ifnum #1<34\Large\else \ifnum #1<41\LARGE\else
     \huge\fi\fi\fi\fi\fi\fi
  \csname #3\endcsname}%
\else
\gdef\SetFigFont#1#2#3{\begingroup
  \count@#1\relax \ifnum 25<\count@\count@25\fi
  \def\x{\endgroup\@setsize\SetFigFont{#2pt}}%
  \expandafter\x
    \csname \romannumeral\the\count@ pt\expandafter\endcsname
    \csname @\romannumeral\the\count@ pt\endcsname
  \csname #3\endcsname}%
\fi
\fi\endgroup
\begin{picture}(324,249)(514,2)
\end{picture}
\right) \right] \times \nn \\
  && \hspace{1.6cm}
      \times \left( \,\,\begin{picture}(0,0)%
\includegraphics{ww14.pstex}%
\end{picture}%
\setlength{\unitlength}{3947sp}%
\begingroup\makeatletter\ifx\SetFigFont\undefined
\def\x#1#2#3#4#5#6#7\relax{\def\x{#1#2#3#4#5#6}}%
\expandafter\x\fmtname xxxxxx\relax \def\y{splain}%
\ifx\x\y   
\gdef\SetFigFont#1#2#3{%
  \ifnum #1<17\tiny\else \ifnum #1<20\small\else
  \ifnum #1<24\normalsize\else \ifnum #1<29\large\else
  \ifnum #1<34\Large\else \ifnum #1<41\LARGE\else
     \huge\fi\fi\fi\fi\fi\fi
  \csname #3\endcsname}%
\else
\gdef\SetFigFont#1#2#3{\begingroup
  \count@#1\relax \ifnum 25<\count@\count@25\fi
  \def\x{\endgroup\@setsize\SetFigFont{#2pt}}%
  \expandafter\x
    \csname \romannumeral\the\count@ pt\expandafter\endcsname
    \csname @\romannumeral\the\count@ pt\endcsname
  \csname #3\endcsname}%
\fi
\fi\endgroup
\begin{picture}(324,324)(889,-73)
\end{picture}
 
      + \,\begin{picture}(0,0)%
\includegraphics{ww23.pstex}%
\end{picture}%
\setlength{\unitlength}{3947sp}%
\begingroup\makeatletter\ifx\SetFigFont\undefined
\def\x#1#2#3#4#5#6#7\relax{\def\x{#1#2#3#4#5#6}}%
\expandafter\x\fmtname xxxxxx\relax \def\y{splain}%
\ifx\x\y   
\gdef\SetFigFont#1#2#3{%
  \ifnum #1<17\tiny\else \ifnum #1<20\small\else
  \ifnum #1<24\normalsize\else \ifnum #1<29\large\else
  \ifnum #1<34\Large\else \ifnum #1<41\LARGE\else
     \huge\fi\fi\fi\fi\fi\fi
  \csname #3\endcsname}%
\else
\gdef\SetFigFont#1#2#3{\begingroup
  \count@#1\relax \ifnum 25<\count@\count@25\fi
  \def\x{\endgroup\@setsize\SetFigFont{#2pt}}%
  \expandafter\x
    \csname \romannumeral\the\count@ pt\expandafter\endcsname
    \csname @\romannumeral\the\count@ pt\endcsname
  \csname #3\endcsname}%
\fi
\fi\endgroup
\begin{picture}(324,324)(814,-73)
\end{picture}
\, \right)  \bigg\} \nn \\
&& + \fr{g^4}{4} 
       \bigg\{ \left[ D_2\left(\,\begin{picture}(0,0)%
\includegraphics{arg3_12.pstex}%
\end{picture}%
\setlength{\unitlength}{3947sp}%
\begingroup\makeatletter\ifx\SetFigFont\undefined
\def\x#1#2#3#4#5#6#7\relax{\def\x{#1#2#3#4#5#6}}%
\expandafter\x\fmtname xxxxxx\relax \def\y{splain}%
\ifx\x\y   
\gdef\SetFigFont#1#2#3{%
  \ifnum #1<17\tiny\else \ifnum #1<20\small\else
  \ifnum #1<24\normalsize\else \ifnum #1<29\large\else
  \ifnum #1<34\Large\else \ifnum #1<41\LARGE\else
     \huge\fi\fi\fi\fi\fi\fi
  \csname #3\endcsname}%
\else
\gdef\SetFigFont#1#2#3{\begingroup
  \count@#1\relax \ifnum 25<\count@\count@25\fi
  \def\x{\endgroup\@setsize\SetFigFont{#2pt}}%
  \expandafter\x
    \csname \romannumeral\the\count@ pt\expandafter\endcsname
    \csname @\romannumeral\the\count@ pt\endcsname
  \csname #3\endcsname}%
\fi
\fi\endgroup
\begin{picture}(249,249)(514,2)
\end{picture}
\right) 
  - D_2\left(\,\begin{picture}(0,0)%
\includegraphics{arg3_13.pstex}%
\end{picture}%
\setlength{\unitlength}{3947sp}%
\begingroup\makeatletter\ifx\SetFigFont\undefined
\def\x#1#2#3#4#5#6#7\relax{\def\x{#1#2#3#4#5#6}}%
\expandafter\x\fmtname xxxxxx\relax \def\y{splain}%
\ifx\x\y   
\gdef\SetFigFont#1#2#3{%
  \ifnum #1<17\tiny\else \ifnum #1<20\small\else
  \ifnum #1<24\normalsize\else \ifnum #1<29\large\else
  \ifnum #1<34\Large\else \ifnum #1<41\LARGE\else
     \huge\fi\fi\fi\fi\fi\fi
  \csname #3\endcsname}%
\else
\gdef\SetFigFont#1#2#3{\begingroup
  \count@#1\relax \ifnum 25<\count@\count@25\fi
  \def\x{\endgroup\@setsize\SetFigFont{#2pt}}%
  \expandafter\x
    \csname \romannumeral\the\count@ pt\expandafter\endcsname
    \csname @\romannumeral\the\count@ pt\endcsname
  \csname #3\endcsname}%
\fi
\fi\endgroup
\begin{picture}(324,249)(514,2)
\end{picture}
\right) 
  + D_2\left(\,\begin{picture}(0,0)%
\includegraphics{arg3_23.pstex}%
\end{picture}%
\setlength{\unitlength}{3947sp}%
\begingroup\makeatletter\ifx\SetFigFont\undefined
\def\x#1#2#3#4#5#6#7\relax{\def\x{#1#2#3#4#5#6}}%
\expandafter\x\fmtname xxxxxx\relax \def\y{splain}%
\ifx\x\y   
\gdef\SetFigFont#1#2#3{%
  \ifnum #1<17\tiny\else \ifnum #1<20\small\else
  \ifnum #1<24\normalsize\else \ifnum #1<29\large\else
  \ifnum #1<34\Large\else \ifnum #1<41\LARGE\else
     \huge\fi\fi\fi\fi\fi\fi
  \csname #3\endcsname}%
\else
\gdef\SetFigFont#1#2#3{\begingroup
  \count@#1\relax \ifnum 25<\count@\count@25\fi
  \def\x{\endgroup\@setsize\SetFigFont{#2pt}}%
  \expandafter\x
    \csname \romannumeral\the\count@ pt\expandafter\endcsname
    \csname @\romannumeral\the\count@ pt\endcsname
  \csname #3\endcsname}%
\fi
\fi\endgroup
\begin{picture}(249,249)(439,2)
\end{picture}
\right) \right]  \times \nn \\
 && \hspace{1.3cm} 
   \times \left(\,\,\begin{picture}(0,0)%
\includegraphics{ww3a.pstex}%
\end{picture}%
\setlength{\unitlength}{3947sp}%
\begingroup\makeatletter\ifx\SetFigFont\undefined
\def\x#1#2#3#4#5#6#7\relax{\def\x{#1#2#3#4#5#6}}%
\expandafter\x\fmtname xxxxxx\relax \def\y{splain}%
\ifx\x\y   
\gdef\SetFigFont#1#2#3{%
  \ifnum #1<17\tiny\else \ifnum #1<20\small\else
  \ifnum #1<24\normalsize\else \ifnum #1<29\large\else
  \ifnum #1<34\Large\else \ifnum #1<41\LARGE\else
     \huge\fi\fi\fi\fi\fi\fi
  \csname #3\endcsname}%
\else
\gdef\SetFigFont#1#2#3{\begingroup
  \count@#1\relax \ifnum 25<\count@\count@25\fi
  \def\x{\endgroup\@setsize\SetFigFont{#2pt}}%
  \expandafter\x
    \csname \romannumeral\the\count@ pt\expandafter\endcsname
    \csname @\romannumeral\the\count@ pt\endcsname
  \csname #3\endcsname}%
\fi
\fi\endgroup
\begin{picture}(324,324)(814,-148)
\end{picture}

   - \,\begin{picture}(0,0)%
\includegraphics{ww3b.pstex}%
\end{picture}%
\setlength{\unitlength}{3947sp}%
\begingroup\makeatletter\ifx\SetFigFont\undefined
\def\x#1#2#3#4#5#6#7\relax{\def\x{#1#2#3#4#5#6}}%
\expandafter\x\fmtname xxxxxx\relax \def\y{splain}%
\ifx\x\y   
\gdef\SetFigFont#1#2#3{%
  \ifnum #1<17\tiny\else \ifnum #1<20\small\else
  \ifnum #1<24\normalsize\else \ifnum #1<29\large\else
  \ifnum #1<34\Large\else \ifnum #1<41\LARGE\else
     \huge\fi\fi\fi\fi\fi\fi
  \csname #3\endcsname}%
\else
\gdef\SetFigFont#1#2#3{\begingroup
  \count@#1\relax \ifnum 25<\count@\count@25\fi
  \def\x{\endgroup\@setsize\SetFigFont{#2pt}}%
  \expandafter\x
    \csname \romannumeral\the\count@ pt\expandafter\endcsname
    \csname @\romannumeral\the\count@ pt\endcsname
  \csname #3\endcsname}%
\fi
\fi\endgroup
\begin{picture}(399,399)(739,-148)
\end{picture}
 
   - \,\begin{picture}(0,0)%
\includegraphics{ww3c.pstex}%
\end{picture}%
\setlength{\unitlength}{3947sp}%
\begingroup\makeatletter\ifx\SetFigFont\undefined
\def\x#1#2#3#4#5#6#7\relax{\def\x{#1#2#3#4#5#6}}%
\expandafter\x\fmtname xxxxxx\relax \def\y{splain}%
\ifx\x\y   
\gdef\SetFigFont#1#2#3{%
  \ifnum #1<17\tiny\else \ifnum #1<20\small\else
  \ifnum #1<24\normalsize\else \ifnum #1<29\large\else
  \ifnum #1<34\Large\else \ifnum #1<41\LARGE\else
     \huge\fi\fi\fi\fi\fi\fi
  \csname #3\endcsname}%
\else
\gdef\SetFigFont#1#2#3{\begingroup
  \count@#1\relax \ifnum 25<\count@\count@25\fi
  \def\x{\endgroup\@setsize\SetFigFont{#2pt}}%
  \expandafter\x
    \csname \romannumeral\the\count@ pt\expandafter\endcsname
    \csname @\romannumeral\the\count@ pt\endcsname
  \csname #3\endcsname}%
\fi
\fi\endgroup
\begin{picture}(324,324)(814,-148)
\end{picture}

   + \,\begin{picture}(0,0)%
\includegraphics{ww3d.pstex}%
\end{picture}%
\setlength{\unitlength}{3947sp}%
\begingroup\makeatletter\ifx\SetFigFont\undefined
\def\x#1#2#3#4#5#6#7\relax{\def\x{#1#2#3#4#5#6}}%
\expandafter\x\fmtname xxxxxx\relax \def\y{splain}%
\ifx\x\y   
\gdef\SetFigFont#1#2#3{%
  \ifnum #1<17\tiny\else \ifnum #1<20\small\else
  \ifnum #1<24\normalsize\else \ifnum #1<29\large\else
  \ifnum #1<34\Large\else \ifnum #1<41\LARGE\else
     \huge\fi\fi\fi\fi\fi\fi
  \csname #3\endcsname}%
\else
\gdef\SetFigFont#1#2#3{\begingroup
  \count@#1\relax \ifnum 25<\count@\count@25\fi
  \def\x{\endgroup\@setsize\SetFigFont{#2pt}}%
  \expandafter\x
    \csname \romannumeral\the\count@ pt\expandafter\endcsname
    \csname @\romannumeral\the\count@ pt\endcsname
  \csname #3\endcsname}%
\fi
\fi\endgroup
\begin{picture}(399,402)(739,-151)
\end{picture}

   \right. \nn \\
  && \left.  \hspace{2.0cm}
   - \,\,\begin{picture}(0,0)%
\includegraphics{ww3e.pstex}%
\end{picture}%
\setlength{\unitlength}{3947sp}%
\begingroup\makeatletter\ifx\SetFigFont\undefined
\def\x#1#2#3#4#5#6#7\relax{\def\x{#1#2#3#4#5#6}}%
\expandafter\x\fmtname xxxxxx\relax \def\y{splain}%
\ifx\x\y   
\gdef\SetFigFont#1#2#3{%
  \ifnum #1<17\tiny\else \ifnum #1<20\small\else
  \ifnum #1<24\normalsize\else \ifnum #1<29\large\else
  \ifnum #1<34\Large\else \ifnum #1<41\LARGE\else
     \huge\fi\fi\fi\fi\fi\fi
  \csname #3\endcsname}%
\else
\gdef\SetFigFont#1#2#3{\begingroup
  \count@#1\relax \ifnum 25<\count@\count@25\fi
  \def\x{\endgroup\@setsize\SetFigFont{#2pt}}%
  \expandafter\x
    \csname \romannumeral\the\count@ pt\expandafter\endcsname
    \csname @\romannumeral\the\count@ pt\endcsname
  \csname #3\endcsname}%
\fi
\fi\endgroup
\begin{picture}(324,324)(514,-148)
\end{picture}

   + \,\begin{picture}(0,0)%
\includegraphics{ww3f.pstex}%
\end{picture}%
\setlength{\unitlength}{3947sp}%
\begingroup\makeatletter\ifx\SetFigFont\undefined
\def\x#1#2#3#4#5#6#7\relax{\def\x{#1#2#3#4#5#6}}%
\expandafter\x\fmtname xxxxxx\relax \def\y{splain}%
\ifx\x\y   
\gdef\SetFigFont#1#2#3{%
  \ifnum #1<17\tiny\else \ifnum #1<20\small\else
  \ifnum #1<24\normalsize\else \ifnum #1<29\large\else
  \ifnum #1<34\Large\else \ifnum #1<41\LARGE\else
     \huge\fi\fi\fi\fi\fi\fi
  \csname #3\endcsname}%
\else
\gdef\SetFigFont#1#2#3{\begingroup
  \count@#1\relax \ifnum 25<\count@\count@25\fi
  \def\x{\endgroup\@setsize\SetFigFont{#2pt}}%
  \expandafter\x
    \csname \romannumeral\the\count@ pt\expandafter\endcsname
    \csname @\romannumeral\the\count@ pt\endcsname
  \csname #3\endcsname}%
\fi
\fi\endgroup
\begin{picture}(399,402)(364,-151)
\end{picture}

   + \,\begin{picture}(0,0)%
\includegraphics{ww3g.pstex}%
\end{picture}%
\setlength{\unitlength}{3947sp}%
\begingroup\makeatletter\ifx\SetFigFont\undefined
\def\x#1#2#3#4#5#6#7\relax{\def\x{#1#2#3#4#5#6}}%
\expandafter\x\fmtname xxxxxx\relax \def\y{splain}%
\ifx\x\y   
\gdef\SetFigFont#1#2#3{%
  \ifnum #1<17\tiny\else \ifnum #1<20\small\else
  \ifnum #1<24\normalsize\else \ifnum #1<29\large\else
  \ifnum #1<34\Large\else \ifnum #1<41\LARGE\else
     \huge\fi\fi\fi\fi\fi\fi
  \csname #3\endcsname}%
\else
\gdef\SetFigFont#1#2#3{\begingroup
  \count@#1\relax \ifnum 25<\count@\count@25\fi
  \def\x{\endgroup\@setsize\SetFigFont{#2pt}}%
  \expandafter\x
    \csname \romannumeral\the\count@ pt\expandafter\endcsname
    \csname @\romannumeral\the\count@ pt\endcsname
  \csname #3\endcsname}%
\fi
\fi\endgroup
\begin{picture}(324,324)(514,-148)
\end{picture}

   - \,\begin{picture}(0,0)%
\includegraphics{ww3h.pstex}%
\end{picture}%
\setlength{\unitlength}{3947sp}%
\begingroup\makeatletter\ifx\SetFigFont\undefined
\def\x#1#2#3#4#5#6#7\relax{\def\x{#1#2#3#4#5#6}}%
\expandafter\x\fmtname xxxxxx\relax \def\y{splain}%
\ifx\x\y   
\gdef\SetFigFont#1#2#3{%
  \ifnum #1<17\tiny\else \ifnum #1<20\small\else
  \ifnum #1<24\normalsize\else \ifnum #1<29\large\else
  \ifnum #1<34\Large\else \ifnum #1<41\LARGE\else
     \huge\fi\fi\fi\fi\fi\fi
  \csname #3\endcsname}%
\else
\gdef\SetFigFont#1#2#3{\begingroup
  \count@#1\relax \ifnum 25<\count@\count@25\fi
  \def\x{\endgroup\@setsize\SetFigFont{#2pt}}%
  \expandafter\x
    \csname \romannumeral\the\count@ pt\expandafter\endcsname
    \csname @\romannumeral\the\count@ pt\endcsname
  \csname #3\endcsname}%
\fi
\fi\endgroup
\begin{picture}(399,399)(364,-148)
\end{picture}
\,
 \right) \bigg\} \nn \\
&& + \fr{g^4}{2} 
   \, D_2\left(\,\begin{picture}(0,0)%
\includegraphics{arg2.pstex}%
\end{picture}%
\setlength{\unitlength}{3947sp}%
\begingroup\makeatletter\ifx\SetFigFont\undefined
\def\x#1#2#3#4#5#6#7\relax{\def\x{#1#2#3#4#5#6}}%
\expandafter\x\fmtname xxxxxx\relax \def\y{splain}%
\ifx\x\y   
\gdef\SetFigFont#1#2#3{%
  \ifnum #1<17\tiny\else \ifnum #1<20\small\else
  \ifnum #1<24\normalsize\else \ifnum #1<29\large\else
  \ifnum #1<34\Large\else \ifnum #1<41\LARGE\else
     \huge\fi\fi\fi\fi\fi\fi
  \csname #3\endcsname}%
\else
\gdef\SetFigFont#1#2#3{\begingroup
  \count@#1\relax \ifnum 25<\count@\count@25\fi
  \def\x{\endgroup\@setsize\SetFigFont{#2pt}}%
  \expandafter\x
    \csname \romannumeral\the\count@ pt\expandafter\endcsname
    \csname @\romannumeral\the\count@ pt\endcsname
  \csname #3\endcsname}%
\fi
\fi\endgroup
\begin{picture}(174,249)(589,2)
\end{picture}
\right) 
   \times \left( 
   \,\,\begin{picture}(0,0)%
\includegraphics{ww2a.pstex}%
\end{picture}%
\setlength{\unitlength}{3947sp}%
\begingroup\makeatletter\ifx\SetFigFont\undefined
\def\x#1#2#3#4#5#6#7\relax{\def\x{#1#2#3#4#5#6}}%
\expandafter\x\fmtname xxxxxx\relax \def\y{splain}%
\ifx\x\y   
\gdef\SetFigFont#1#2#3{%
  \ifnum #1<17\tiny\else \ifnum #1<20\small\else
  \ifnum #1<24\normalsize\else \ifnum #1<29\large\else
  \ifnum #1<34\Large\else \ifnum #1<41\LARGE\else
     \huge\fi\fi\fi\fi\fi\fi
  \csname #3\endcsname}%
\else
\gdef\SetFigFont#1#2#3{\begingroup
  \count@#1\relax \ifnum 25<\count@\count@25\fi
  \def\x{\endgroup\@setsize\SetFigFont{#2pt}}%
  \expandafter\x
    \csname \romannumeral\the\count@ pt\expandafter\endcsname
    \csname @\romannumeral\the\count@ pt\endcsname
  \csname #3\endcsname}%
\fi
\fi\endgroup
\begin{picture}(399,324)(814,-148)
\end{picture}

   - \,\begin{picture}(0,0)%
\includegraphics{ww2b.pstex}%
\end{picture}%
\setlength{\unitlength}{3947sp}%
\begingroup\makeatletter\ifx\SetFigFont\undefined
\def\x#1#2#3#4#5#6#7\relax{\def\x{#1#2#3#4#5#6}}%
\expandafter\x\fmtname xxxxxx\relax \def\y{splain}%
\ifx\x\y   
\gdef\SetFigFont#1#2#3{%
  \ifnum #1<17\tiny\else \ifnum #1<20\small\else
  \ifnum #1<24\normalsize\else \ifnum #1<29\large\else
  \ifnum #1<34\Large\else \ifnum #1<41\LARGE\else
     \huge\fi\fi\fi\fi\fi\fi
  \csname #3\endcsname}%
\else
\gdef\SetFigFont#1#2#3{\begingroup
  \count@#1\relax \ifnum 25<\count@\count@25\fi
  \def\x{\endgroup\@setsize\SetFigFont{#2pt}}%
  \expandafter\x
    \csname \romannumeral\the\count@ pt\expandafter\endcsname
    \csname @\romannumeral\the\count@ pt\endcsname
  \csname #3\endcsname}%
\fi
\fi\endgroup
\begin{picture}(399,474)(739,-223)
\end{picture}

   - \,\begin{picture}(0,0)%
\includegraphics{ww2c.pstex}%
\end{picture}%
\setlength{\unitlength}{3947sp}%
\begingroup\makeatletter\ifx\SetFigFont\undefined
\def\x#1#2#3#4#5#6#7\relax{\def\x{#1#2#3#4#5#6}}%
\expandafter\x\fmtname xxxxxx\relax \def\y{splain}%
\ifx\x\y   
\gdef\SetFigFont#1#2#3{%
  \ifnum #1<17\tiny\else \ifnum #1<20\small\else
  \ifnum #1<24\normalsize\else \ifnum #1<29\large\else
  \ifnum #1<34\Large\else \ifnum #1<41\LARGE\else
     \huge\fi\fi\fi\fi\fi\fi
  \csname #3\endcsname}%
\else
\gdef\SetFigFont#1#2#3{\begingroup
  \count@#1\relax \ifnum 25<\count@\count@25\fi
  \def\x{\endgroup\@setsize\SetFigFont{#2pt}}%
  \expandafter\x
    \csname \romannumeral\the\count@ pt\expandafter\endcsname
    \csname @\romannumeral\the\count@ pt\endcsname
  \csname #3\endcsname}%
\fi
\fi\endgroup
\begin{picture}(399,474)(364,-223)
\end{picture}

   + \,\begin{picture}(0,0)%
\includegraphics{ww2d.pstex}%
\end{picture}%
\setlength{\unitlength}{3947sp}%
\begingroup\makeatletter\ifx\SetFigFont\undefined
\def\x#1#2#3#4#5#6#7\relax{\def\x{#1#2#3#4#5#6}}%
\expandafter\x\fmtname xxxxxx\relax \def\y{splain}%
\ifx\x\y   
\gdef\SetFigFont#1#2#3{%
  \ifnum #1<17\tiny\else \ifnum #1<20\small\else
  \ifnum #1<24\normalsize\else \ifnum #1<29\large\else
  \ifnum #1<34\Large\else \ifnum #1<41\LARGE\else
     \huge\fi\fi\fi\fi\fi\fi
  \csname #3\endcsname}%
\else
\gdef\SetFigFont#1#2#3{\begingroup
  \count@#1\relax \ifnum 25<\count@\count@25\fi
  \def\x{\endgroup\@setsize\SetFigFont{#2pt}}%
  \expandafter\x
    \csname \romannumeral\the\count@ pt\expandafter\endcsname
    \csname @\romannumeral\the\count@ pt\endcsname
  \csname #3\endcsname}%
\fi
\fi\endgroup
\begin{picture}(474,399)(739,-148)
\end{picture}
 \,
   \right) 
\,.
\label{neudarstdiag}
\eea
Here an integration with the weight $1/(2\pi)^3$ over the loop momentum 
is implied as are the $\delta$-functions according to the gluon lines that 
are not involved in the interactions. 

The different sums of $D_2$'s in the 
new representation (\ref{neudarst}) 
are familiar objects: The combinations in the first integral, 
consisting of three and four $D_2$'s which occur together with 
their sum (seven $D_2$'s), are the momentum space 
factors corresponding to two different color structures of 
the reggeizing part $D_4^R$ in the four--gluon amplitude. 
The combination of three $D_2$'s in the second integral 
is the momentum part of the three--gluon amplitude $D_3$. 
Together with the above discussion of the kernels $K_{2\rarr m}$ 
this emphasizes that the new representation is closely related to the 
elements entering the original integral equation (\ref{inteq4}). 

Interestingly, also the function $G$ (see eq.\ (\ref{gdef})) can be 
expressed through BFKL kernels, 
\bea
\label{gneudarst}
G(\kf_1,\kf_2,\kf_3) &=& 
\frac{g^3}{(2\pi)^3} \int 
\left( \prod_{i=1}^{2} d^2\lf_i \right)
\delta \left( \sum_{j=1}^2 \lf_j - \sum_{j=1}^3 \kf_j \right) 
D_2(\lf_1,\lf_2) 
\times
\nn \\
&& \hspace{0.3cm}
\times \left[ {\cal K}(\lf_1,\lf_2;\kf_1+\kf_3,\kf_2)
- {\cal K}(\lf_1-\kf_1,\lf_2;\kf_2,\kf_3)
\right.
\nn \\ 
&& \left.
\hspace{0.8cm}
- {\cal K}(\lf_1,\lf_2-\kf_3;\kf_1,\kf_2)
\right]
\nn \\
&& + \frac{g^3}{2 (2\pi)^3} \int 
\left( \prod_{i=1}^{3} d^2\lf_i \right)
\delta \left( \sum_{j=1}^3 \lf_j - \sum_{j=1}^3 \kf_j \right) 
\times
\nn \\
&& \hspace{0.3cm}
\times 
\left[D_2(\lf_1+\lf_2,\lf_3) - D_2(\lf_1+\lf_3,\lf_2) 
+ D_2(\lf_1,\lf_2+\lf_3) \right] \times
\nn \\
&& \hspace{0.3cm}
\times
\left[ {\cal K}(\lf_1,\lf_2;\kf_1,\kf_2) \delta(\lf_3-\kf_3) 
+ {\cal K}(\lf_1,\lf_3;\kf_1,\kf_3) \delta(\lf_2-\kf_2)
\right.
\nn \\
&& \hspace{0.8cm}
\left.
+ {\cal K}(\lf_2,\lf_3;\kf_2,\kf_3) \delta(\lf_1-\kf_1)
\right]
\,.
\eea
To make this easier to read, we again write it using the 
graphical notation introduced above:
\bea
\label{gneudarstdiag}
G(\kf_1,\kf_2,\kf_3) &=& g^3
D_2\left(\, \begin{picture}(0,0)%
\includegraphics{arg2.pstex}%
\end{picture}%
\setlength{\unitlength}{3947sp}%
\begingroup\makeatletter\ifx\SetFigFont\undefined
\def\x#1#2#3#4#5#6#7\relax{\def\x{#1#2#3#4#5#6}}%
\expandafter\x\fmtname xxxxxx\relax \def\y{splain}%
\ifx\x\y   
\gdef\SetFigFont#1#2#3{%
  \ifnum #1<17\tiny\else \ifnum #1<20\small\else
  \ifnum #1<24\normalsize\else \ifnum #1<29\large\else
  \ifnum #1<34\Large\else \ifnum #1<41\LARGE\else
     \huge\fi\fi\fi\fi\fi\fi
  \csname #3\endcsname}%
\else
\gdef\SetFigFont#1#2#3{\begingroup
  \count@#1\relax \ifnum 25<\count@\count@25\fi
  \def\x{\endgroup\@setsize\SetFigFont{#2pt}}%
  \expandafter\x
    \csname \romannumeral\the\count@ pt\expandafter\endcsname
    \csname @\romannumeral\the\count@ pt\endcsname
  \csname #3\endcsname}%
\fi
\fi\endgroup
\begin{picture}(174,249)(589,2)
\end{picture}
\right) 
\times 
\left( \,\, \begin{picture}(0,0)%
\epsfig{file=gwwa.pstex}%
\end{picture}%
\setlength{\unitlength}{4144sp}%
\begingroup\makeatletter\ifx\SetFigFont\undefined%
\gdef\SetFigFont#1#2#3#4#5{%
  \reset@font\fontsize{#1}{#2pt}%
  \fontfamily{#3}\fontseries{#4}\fontshape{#5}%
  \selectfont}%
\fi\endgroup%
\begin{picture}(339,294)(1249,-73)
\end{picture}
 
- \,\begin{picture}(0,0)%
\epsfig{file=gwwb.pstex}%
\end{picture}%
\setlength{\unitlength}{4144sp}%
\begingroup\makeatletter\ifx\SetFigFont\undefined%
\gdef\SetFigFont#1#2#3#4#5{%
  \reset@font\fontsize{#1}{#2pt}%
  \fontfamily{#3}\fontseries{#4}\fontshape{#5}%
  \selectfont}%
\fi\endgroup%
\begin{picture}(310,381)(703,-101)
\end{picture}
 
- \, \begin{picture}(0,0)%
\epsfig{file=gwwc.pstex}%
\end{picture}%
\setlength{\unitlength}{4144sp}%
\begingroup\makeatletter\ifx\SetFigFont\undefined%
\gdef\SetFigFont#1#2#3#4#5{%
  \reset@font\fontsize{#1}{#2pt}%
  \fontfamily{#3}\fontseries{#4}\fontshape{#5}%
  \selectfont}%
\fi\endgroup%
\begin{picture}(310,381)(417,-101)
\end{picture}
 \,
\right)
\nn \\
&&  + \frac{g^3}{2}\left[ D_2\left(\,\begin{picture}(0,0)%
\includegraphics{arg3_12.pstex}%
\end{picture}%
\setlength{\unitlength}{3947sp}%
\begingroup\makeatletter\ifx\SetFigFont\undefined
\def\x#1#2#3#4#5#6#7\relax{\def\x{#1#2#3#4#5#6}}%
\expandafter\x\fmtname xxxxxx\relax \def\y{splain}%
\ifx\x\y   
\gdef\SetFigFont#1#2#3{%
  \ifnum #1<17\tiny\else \ifnum #1<20\small\else
  \ifnum #1<24\normalsize\else \ifnum #1<29\large\else
  \ifnum #1<34\Large\else \ifnum #1<41\LARGE\else
     \huge\fi\fi\fi\fi\fi\fi
  \csname #3\endcsname}%
\else
\gdef\SetFigFont#1#2#3{\begingroup
  \count@#1\relax \ifnum 25<\count@\count@25\fi
  \def\x{\endgroup\@setsize\SetFigFont{#2pt}}%
  \expandafter\x
    \csname \romannumeral\the\count@ pt\expandafter\endcsname
    \csname @\romannumeral\the\count@ pt\endcsname
  \csname #3\endcsname}%
\fi
\fi\endgroup
\begin{picture}(249,249)(514,2)
\end{picture}
\right) 
  - D_2\left(\,\begin{picture}(0,0)%
\includegraphics{arg3_13.pstex}%
\end{picture}%
\setlength{\unitlength}{3947sp}%
\begingroup\makeatletter\ifx\SetFigFont\undefined
\def\x#1#2#3#4#5#6#7\relax{\def\x{#1#2#3#4#5#6}}%
\expandafter\x\fmtname xxxxxx\relax \def\y{splain}%
\ifx\x\y   
\gdef\SetFigFont#1#2#3{%
  \ifnum #1<17\tiny\else \ifnum #1<20\small\else
  \ifnum #1<24\normalsize\else \ifnum #1<29\large\else
  \ifnum #1<34\Large\else \ifnum #1<41\LARGE\else
     \huge\fi\fi\fi\fi\fi\fi
  \csname #3\endcsname}%
\else
\gdef\SetFigFont#1#2#3{\begingroup
  \count@#1\relax \ifnum 25<\count@\count@25\fi
  \def\x{\endgroup\@setsize\SetFigFont{#2pt}}%
  \expandafter\x
    \csname \romannumeral\the\count@ pt\expandafter\endcsname
    \csname @\romannumeral\the\count@ pt\endcsname
  \csname #3\endcsname}%
\fi
\fi\endgroup
\begin{picture}(324,249)(514,2)
\end{picture}
\right) 
  + D_2\left(\,\begin{picture}(0,0)%
\includegraphics{arg3_23.pstex}%
\end{picture}%
\setlength{\unitlength}{3947sp}%
\begingroup\makeatletter\ifx\SetFigFont\undefined
\def\x#1#2#3#4#5#6#7\relax{\def\x{#1#2#3#4#5#6}}%
\expandafter\x\fmtname xxxxxx\relax \def\y{splain}%
\ifx\x\y   
\gdef\SetFigFont#1#2#3{%
  \ifnum #1<17\tiny\else \ifnum #1<20\small\else
  \ifnum #1<24\normalsize\else \ifnum #1<29\large\else
  \ifnum #1<34\Large\else \ifnum #1<41\LARGE\else
     \huge\fi\fi\fi\fi\fi\fi
  \csname #3\endcsname}%
\else
\gdef\SetFigFont#1#2#3{\begingroup
  \count@#1\relax \ifnum 25<\count@\count@25\fi
  \def\x{\endgroup\@setsize\SetFigFont{#2pt}}%
  \expandafter\x
    \csname \romannumeral\the\count@ pt\expandafter\endcsname
    \csname @\romannumeral\the\count@ pt\endcsname
  \csname #3\endcsname}%
\fi
\fi\endgroup
\begin{picture}(249,249)(439,2)
\end{picture}
\right) \right] \times \nn \\
&& \hspace{0.9cm} \times \left(\,\,\begin{picture}(0,0)%
\epsfig{file=gww12.pstex}%
\end{picture}%
\setlength{\unitlength}{4144sp}%
\begingroup\makeatletter\ifx\SetFigFont\undefined%
\gdef\SetFigFont#1#2#3#4#5{%
  \reset@font\fontsize{#1}{#2pt}%
  \fontfamily{#3}\fontseries{#4}\fontshape{#5}%
  \selectfont}%
\fi\endgroup%
\begin{picture}(238,310)(846,-30)
\end{picture}
 
  + \,\begin{picture}(0,0)%
\epsfig{file=gww13.pstex}%
\end{picture}%
\setlength{\unitlength}{4144sp}%
\begingroup\makeatletter\ifx\SetFigFont\undefined%
\gdef\SetFigFont#1#2#3#4#5{%
  \reset@font\fontsize{#1}{#2pt}%
  \fontfamily{#3}\fontseries{#4}\fontshape{#5}%
  \selectfont}%
\fi\endgroup%
\begin{picture}(238,310)(846,-30)
\end{picture}
 
  + \,\begin{picture}(0,0)%
\epsfig{file=gww23.pstex}%
\end{picture}%
\setlength{\unitlength}{4144sp}%
\begingroup\makeatletter\ifx\SetFigFont\undefined%
\gdef\SetFigFont#1#2#3#4#5{%
  \reset@font\fontsize{#1}{#2pt}%
  \fontfamily{#3}\fontseries{#4}\fontshape{#5}%
  \selectfont}%
\fi\endgroup%
\begin{picture}(239,310)(774,-30)
\end{picture}
 \,\right) 
\,.
\eea
Here we again find the specific combination of three $D_2$ amplitudes 
which is the momentum part of the three-gluon amplitude. 

Concerning the relation of the two new representations 
for $V$ and $G$ we should mention that it is not 
possible to derive the representation (\ref{neudarst}) 
from (\ref{gneudarst}) and (\ref{vmitg}) without decomposing the 
BFKL kernel according to (\ref{Lipatovkernel}). 

\boldmath
\section{Possible applications of the new representation}
\unboldmath
\label{implications}

In the preceding section we have found  
a relation between the two--to--four vertex and the 
much simpler BFKL kernel. Now we turn to possible 
applications of this relation. 

Firstly, the new representation is useful for 
the further investigation of the vertex itself. 
The emergence of conformal invariance in the 
complicated vertex seems much more natural 
in view of the fact that the vertex is build from 
conformally invariant BFKL kernels and free propagators.  
The new representation might also be useful for 
studying the interesting but 
difficult question how the vertex behaves under 
crossing, i.\,e.\ what its properties are under the 
exchange of one of the two incoming and one of the four 
outgoing gluons. 

Unitarity corrections have also been studied in 
Mueller's dipole picture \cite{dipole}. 
Its relation to the field theory structure arising in the 
$t$-channel approach has not yet been clarified beyond 
the one--ladder (BFKL) approximation. We hope that a structure 
similar to our new representation of the vertex can 
also be identified in the dipole picture. 

Recently, the NLO corrections to the BFKL kernel have been 
calculated \cite{FL,CC}. A natural question is 
whether it is possible to find also the NLO corrections 
to the other elements of the effective field theory of unitarity 
corrections, especially the corrections to the two--to--four vertex.
On first sight this appears to be extremely 
difficult, since it would require to derive and solve  
in NLO the full integral equations describing the $n$-gluon 
amplitudes. On the other hand, our new representation 
shows that the vertex is closely related to the BFKL 
kernel. If this relation could be shown to hold in NLO as well, 
then it would be possible to obtain the NLO two--to--four gluon 
vertex simply from eq.\ (\ref{neudarst}) by 
replacing the LO BFKL kernel by its NLO modification 
\be
\label{replace}
{\cal K} \longrightarrow {\cal K}^{\mbox{\scriptsize NLO}} 
\,,
\ee
as calculated in \cite{FL}. The crucial point is that it is 
most probably easier to prove the relation between the vertex 
and the BFKL kernel in NLO than to use the full machinery 
of the integral equations to compute the NLO vertex. 
We expect that the origin of the new representation lies 
in some kind of bootstrap relation for the kernels 
$K_{2\rarr m}$ in the integral equations together with 
the symmetry of the amplitudes in color space. Further, it 
would be necessary to prove the reggeization of certain 
parts of the $n$-gluon amplitudes (i.\,e.\ their being 
superpositions of BFKL amplitudes) in NLO. 
These properties are very general features of QCD in the high 
energy limit and should presumably also hold in NLO, 
although a proof is still missing. 
In the absence of a full computation from first principles, 
the representation (\ref{neudarst}) together with the 
replacement (\ref{replace}) can at least be used as an educated 
guess for the NLO two--to--four gluon vertex. 
This conjectured form can then be tested in view of more 
general properties expected for transition vertices in the 
effective theory of unitarity corrections. 
The conjecture can also be applied to other object of interest 
in high energy QCD. The triple--Pomeron vertex, for example, 
can be obtained from the two--to--four vertex after projection 
onto three BFKL Pomeron states \cite{Hans,Gregory3p}. 
Its NLO corrections can thus also be obtained from the 
NLO vertex. 

Finally, we expect that a representation in terms of BFKL kernels 
can also be constructed for higher transition vertices like the 
two--to--six transition found in \cite{BE}. The BFKL kernel 
thus turns out to play an important role not only in the two--gluon 
state but in the whole set of unitarity corrections. 

\section{Summary}

We have shown that the two--to--four gluon vertex 
can be expressed in terms of BFKL kernels, thus establishing 
the relation between these two elements of the effective 
field theory of unitarity corrections. The underlying principles 
of the effective field theory leading to this relation are still unknown. 
We hope that the new representation can help in finding and 
understanding them. We have indicated possible applications 
of the new representation which deserve further study, 
among them the possibility of finding the NLO corrections 
to the two--to--four gluon vertex.

\section*{Acknowledgements}
I would like to thank Jochen Bartels, Peter Landshoff, 
and Bryan Webber for helpful discussions.

\end{document}